\documentclass[%
reprint,
superscriptaddress,
showpacs,
 amsmath,amssymb,
 prl
]{revtex4-1}

\usepackage{graphicx}
\usepackage{dcolumn}
\usepackage{bm}
\usepackage{xcolor}

\begin{document}

\title{Noise-enhanced spatial-photonic Ising machine}

\author{Davide Pierangeli}
\email{Davide.Pierangeli@roma1.infn.it}
\affiliation{Dipartimento di Fisica, Universit\`{a} di Roma  ``La Sapienza'', 00185 Rome, Italy}
\affiliation{Institute for Complex System, National Research Council (ISC-CNR), 00185 Rome, Italy}

\author{Giulia Marcucci}
\affiliation{Dipartimento di Fisica, Universit\`{a} di Roma  ``La Sapienza'', 00185 Rome, Italy}
\affiliation{Institute for Complex System, National Research Council (ISC-CNR), 00185 Rome, Italy}

\author{Daniel Brunner}
\affiliation{FEMTO-ST Institute/Optics Department CNRS \& Universite Bourgogne Franche-Comte 15B Avenue des Montboucons F-25030 Besanccon, France}

\author{Claudio Conti}
\affiliation{Institute for Complex System, National Research Council (ISC-CNR), 00185 Rome, Italy}
\affiliation{Dipartimento di Fisica, Universit\`{a} di Roma  ``La Sapienza'', 00185 Rome, Italy}

\begin{abstract}
{\bf Abstract:} 
Ising machines are novel computing devices for the energy minimization of Ising models. These combinatorial optimization problems are of paramount importance for science and technology, but remain difficult to tackle on large scale by conventional electronics. Recently, various photonics-based Ising machines demonstrated ultra-fast computing of Ising ground state by data processing through multiple temporal or spatial optical channels. 
Experimental noise acts as a detrimental effect in many of these devices.
On the contrary, we here demonstrate that an optimal noise level enhances the performance of spatial-photonic Ising machines
on frustrated spin problems. By controlling the error rate at the detection, 
we introduce a noisy-feedback mechanism in an Ising machine based on spatial light modulation.
We investigate the device performance on systems with hundreds of individually-addressable spins with all-to-all couplings
and we found an increased success probability at a specific noise level. The optimal noise amplitude depends on graph properties and size, thus indicating an additional tunable parameter helpful in exploring complex energy landscapes and in avoiding trapping into local minima. 
The result points out noise as a resource for optical computing. 
This concept, which also holds in different nanophotonic neural networks, may be crucial in developing novel hardware with optics-enabled parallel
architecture for large-scale optimizations.

\vspace*{+0.2cm}
\end{abstract}

\maketitle

\section{Introduction}

Solving large combinatorial problems is crucial for widespread applications in fields such as artificial intelligence, cryptography, biophysics, and complex networks. However, finding the optimal solution to many of these tasks requires resources that grow exponentially with the problem size, a reason why 
they are considered as computationally intractable for traditional computing architectures \cite{Hromkovic2013}. 
A promising approach to efficiently solve these problems is to recast them in terms of an Ising model \cite{Barahona1982, Lucas2014}, which describes a system of classical interacting spins, and searching its ground state by an artificial network of spins evolving by an Ising Hamiltonian.
Ising machines are physical platforms made of atoms, electrons, or photons, that can be programmed to encode Ising problems with known coupling values and enable to find the optimal solution by the measurement outcome.
They have been realized in a variety of quantum and classical systems including cold atoms \cite{Kim2010, Simon2011}, single photons \cite{Ma2011, Sciarrino2017}, superconducting \cite{Johnson2011, Boixo2014} and magnetic junctions \cite{Datta2019},  electromechanical \cite{Mahboob2016} and CMOS circuits \cite{Yamaoka2016}, polariton and photon condensates \cite{Berloff2017, Dung2017}, or lasers and nonlinear waves \cite{Ghofraniha2015, Nixon2013, Pierangeli2017}, but with practical difficulties in scalability, connectivity, or in engineering the spin interaction.

Photonic Ising machines encode the spin state in the phase or amplitude of the optical field. These devices process data in parallel at high speed through active optical components and can be much faster than those based on other encoding schemes \cite{Brunner2013}.
Various prototypes have recently been realized with sizes spanning from few to thousands of spins. 
In the class of photonic optimizers known as coherent Ising machines (CIMs), the nonlinear dynamics of time-multiplexed optical parametric oscillators \cite{Marandi2014, McMahon2016, Inagaki2016, Inagaki2016_2, Takeda2017}, fiber lasers \cite{Babaeian2019}, or simple opto-electronic oscillators \cite{Vandersande2019}, is exploited to solve NP-hard optimization problems with notable performance~\cite{Hamerly2019_2}.
CIMs are dissipative optical networks in which the ground-state search is performed in reverse direction by slowly raising the gain, according to a general non-equilibrium bifurcation mechanism that currently inspires novel algorithms and settings 
\cite{Goto2019, Kalinin2018, Blais2017, Lvovsky2019, Clements2017, Bohm2018, Kalinin2018_2, Leleu2019,  Bello2019, Chou2019, Davidson2019}.
On the other hand, optimization platforms based on waveguides circuits \cite{Wu2014, Soci2018} and integrated nanophotonic processors \cite{Roques-Carmes2020, Prabhu2019, Shen2017, Harris2018} operate as optical recurrent neural networks \cite{Brunner2018} converging to Ising energy minima. 

Spatial-photonic Ising machines are a different class of optical Ising solvers that have been demonstrated very recently \cite{Pierangeli2019}.
They make use of spatial light modulation for encoding an unprecedented number of spins \cite{Roques-Carmes2019},
while the programmed Hamiltonian is optically evaluated by measuring the intensity distribution after propagation in free-space \cite{Pierangeli2019} or through nonlinear media \cite{Kumar2020}. These devices take advantage of the parallelization provided by spatial multiplexing and by the large pixel density of spatial light modulators (SLMs), thus enabling the implementation of large-scale neuromorphic computing  \cite{Hamerly2019, Zuo2019, Saade2016, Dong2020,  Hougne2018, Popoff2019}.

\begin{figure*}[t!]
\centering
\vspace*{-0.2cm}
\hspace*{-0.3cm} 
\includegraphics[width=1.80\columnwidth]{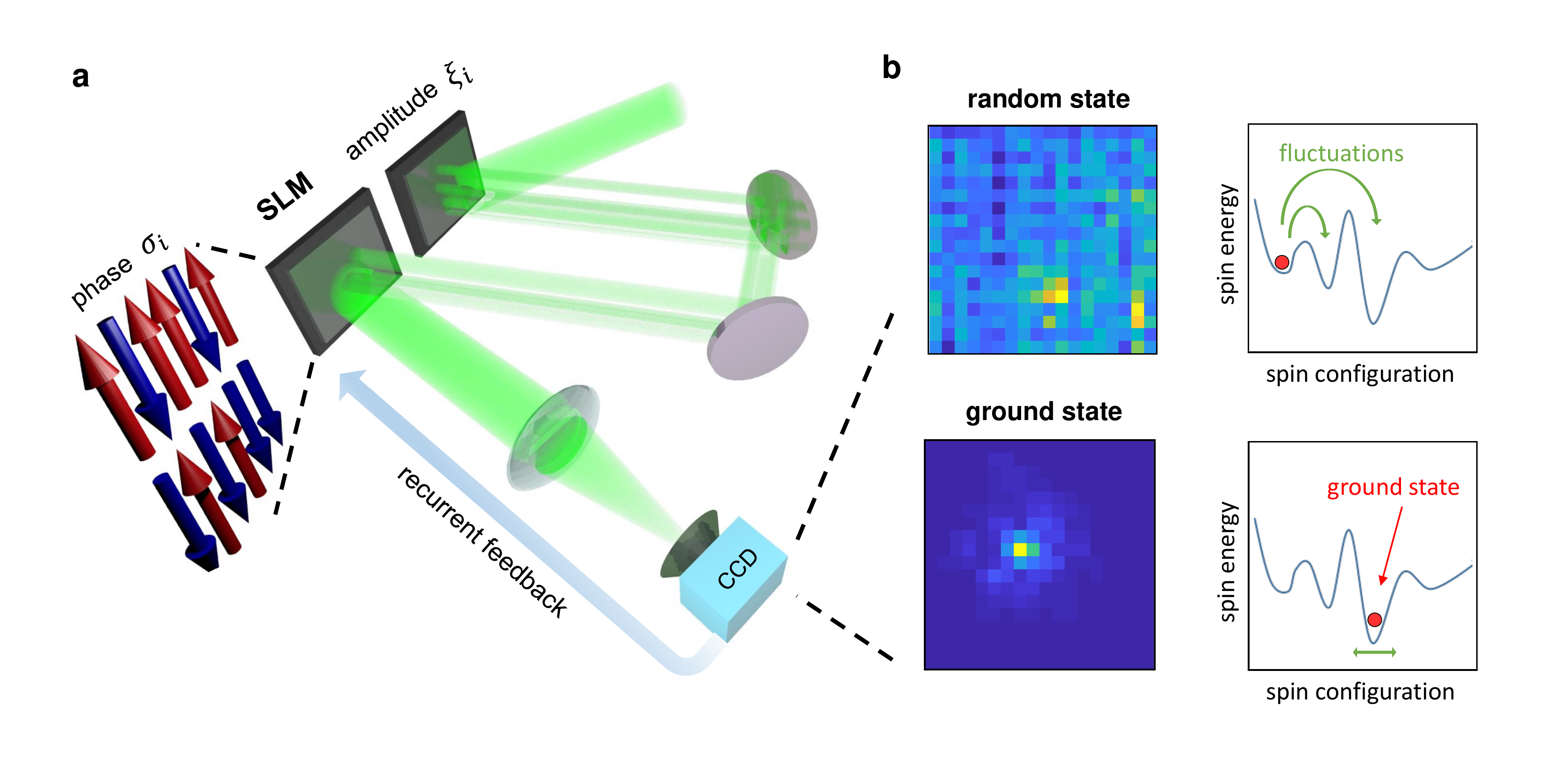} 
\vspace*{-0.7cm}
\caption{{\bf Photonic Ising machine by spatial light modulation.} (a) The spins $\sigma_i = \pm 1$ are encoded by a spatial light modulator (SLM) into binary optical phases $\phi= 0, \pi$ in separated spatial points of the optical wavefront. Intensity modulation is employed
to set the spin interaction via the amplitude distribution $\xi_i$. Recurrent feedback from the far-field camera allows evolution of the phase configuration towards the Ising ground state. (b) Intensity detected on the CCD modes for a random spin state and for the ground state of a ferromagnetic Ising model.  Insets are illustrations of the corresponding spin state in energy landscape. Experimental fluctuations allow spontaneous changes in the spin configuration.
}
\vspace{-0.2cm}
\label{Figure1}
\end{figure*}

Noise is an unavoidable ingredient in any hardware, even more in optical Ising machines. In CIMs it represents one of the main error sources, which can lead the dynamics beyond the regime of Ising spins \cite{Strinati2019}.
On the contrary, in recurrent algorithms and artificial neural networks for Ising problems, noise furnishes a finite effective temperature, and it is expected to facilitate and speed up convergence to the ground state \cite{Roques-Carmes2020, Cai2019}.
The effect has been recently observed in a platform with few photonic spins and various competing interactions \cite{Prabhu2019}.
Noise-tolerant settings are especially important when scaling the device to solve systems with many units.
Nevertheless, the impact of noise in large-scale photonic Ising machines remains mainly unexplored.

In the present article, we investigate the effect of experimental noise on a spatial-photonic Ising machine with hundreds of spins, 
proving the existence of an optimal noise level. The impact of noise depends on the problem features. 
Specifically, we found that it enhances the machine's success probability on problems with both positive and negative interactions, while it is detrimental for models having only positive couplings. 
By providing a mechanism to escape from local minima, noise represents an additional parameter with beneficial properties for our optical computing device, in close analogy with recurrent neural networks trained to solve Ising problems.
Our findings demonstrate noise as a valuable resource in large-scale photonic computing.

\section{Ising machine by spatial light modulation}

In our Ising machine the spin variables are encoded on a coherent laser wavefront via binary values of the optical phase and processed
by spatial light modulation \cite{Pierangeli2019}. As schematized in Fig. 1(a), our optical setting employs an optical path in which a SLM encodes
spins $\sigma_i = \pm 1$ by $ 0,\pi$ phase delays over an amplitude-modulated beam. 
Linear interaction between spins occurs by interference on the detection plane and its strength is controlled by spatial modulation of the input intensity.
The optical machine works in a measurement and feedback scheme. Once initialized to a given problem, feedback from the detected intensity allows the phase distribution on the SLM to converge towards a state minimizing an Ising Hamiltonian
\begin{equation}
H= - \sum_{ij} J_{ij} \sigma_i \sigma_j
\end{equation}
with couplings $J_{ij}=\xi_i \xi_j \tilde{I}_T$, where $\xi_i$ is the amplitude illuminating the the $i$-th spin. 
$\tilde{I}_T$ is the Fourier transform of a pre-determined target image, and the difference between $I_T$ and the image detected on the camera is the cost function. At each machine iteration, we measure the intensity on the CCD modes [Fig. 1(b)] and the spins are updated in order to minimize the cost function. At variance with CIMs \cite{McMahon2016, Inagaki2016, Vandersande2019}, the machine operates without electronically computing neither the energy nor the force acting on each spin. 

\begin{figure*}[t!]
\centering
\vspace*{-0.3cm}
\hspace*{-0.3cm} 
\includegraphics[width=1.80\columnwidth]{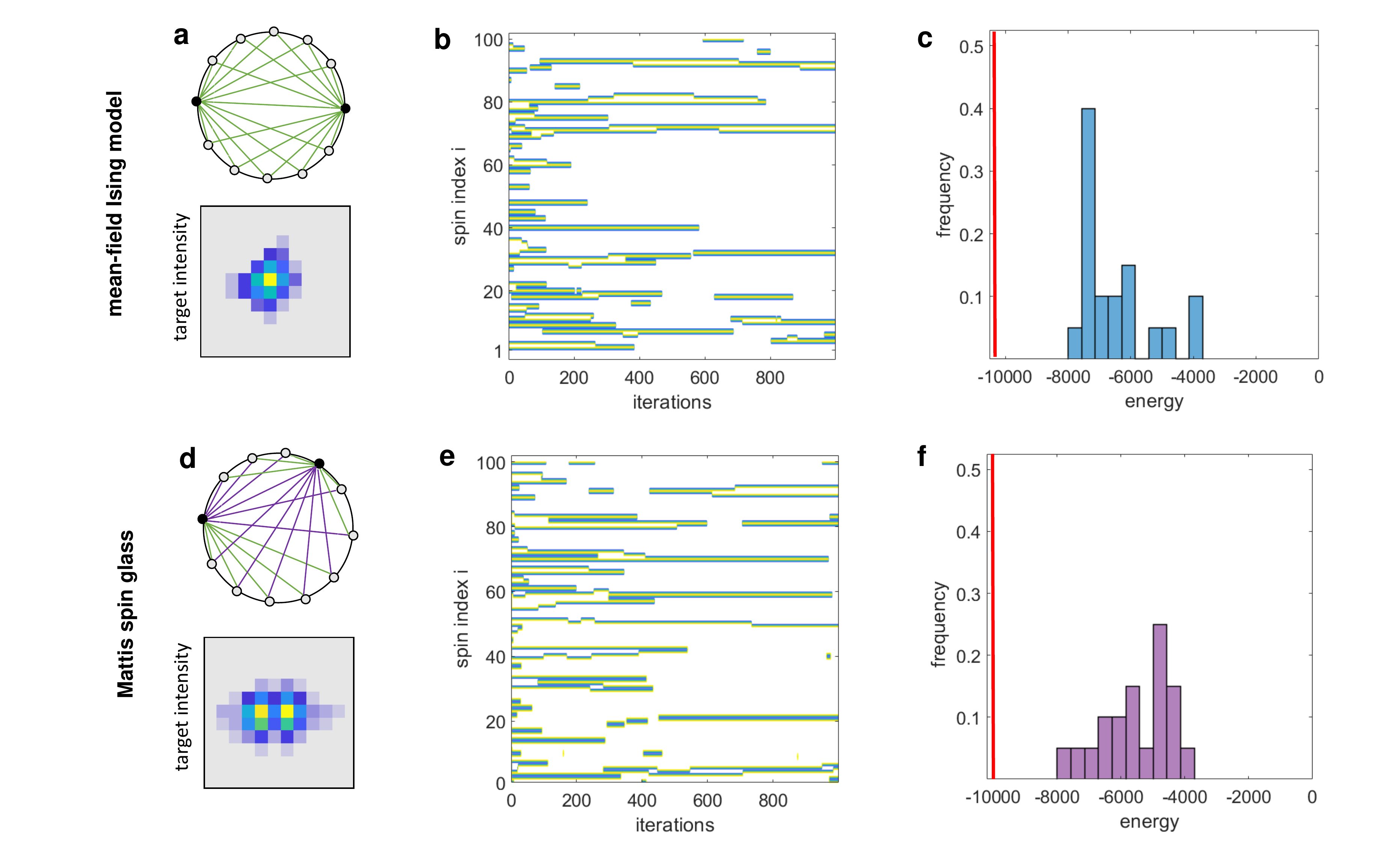} 
\vspace*{-0.2cm}
\caption{ {\bf Ising machine's ground states without noise control.} (a-c) Results for ferromagnetic and (d-f) frustrated models with $100$ spins ($\rho=0$). Problem graphs for (a) mean-field (infinite-range) Ising model and (d) Mattis spin glass with positive and negative couplings. For clarity, only links starting from the two black nodes are drawn. Insets show the corresponding target images $I_T$. 
(b,e) Contour maps showing the evolution of the spin configuration during a single run of the machine and  
(c,f) normalized histograms of the ground-state energy obtained for the instances in (a) and (d), respectively. The red lines in (c) and (f) indicate the minimum energy known from exact solution of the models.
}
\vspace{-0.1cm}
\label{Figure2}
\end{figure*}

Due to the intrinsic noise characterizing each operation in the experimental setup, the machine always behaves as coupled to a thermal bath.
Noise provides an effective temperature for the final spin ground state, as reported in Ref. \cite{Pierangeli2019}.
Sources of noise come from intensity discretization and processing in each CCD mode, as well as from the imperfect spatial light modulation.
As detailed below, we here control the noise level by means of a tunable error rate in the machine's measurement and feedback scheme.

\subsection{Experimental setup and noisy-feedback method}

The experimental device follows the setup illustrated in Fig. 1(a). Light from a CW laser at $\lambda= 532$ nm is expanded and polarization controlled.
The beam is first spatially modulated in amplitude and then in phase by a single reflective modulator (Holoeye LC-R 720, $1280\times768$ pixels, pixel pitch $20\times20~\mu$m). A portion of the modulator operates in amplitude mode to generate the profiles $\xi_i$, which are imaged by a 4-f system on the second SLM portion that perform binary phase modulation. 
Phase modulation occurs with less than 10\% residual intensity modulation.
We select an active area of approximately $200\times200$ SLM pixels and divide it into $N$ addressable optical spins by grouping several pixels.
Modulated light is spatially filtered using an holographic grating and focused by a $f=500$~mm lens on a CCD camera.
The intensity is detected on $18 \times 18$ spatial modes, where the signal in each mode is obtained  averaging over $10 \times 10$ camera pixels, 
a size comparable with the spatial extent of a speckle grain. 

The measured intensity pattern determines the feedback signal. At each machine cycle a single spin is randomly selected and flipped;   
the recorded image is compared with the reference $I_T$ on the same set of modes, and the spin state is updated to minimize the difference between the two images. Due to errors at the readout, always exists a finite probability to update the spin configuration.
Readout errors fixes the noise level and are related to the magnitude of the intensity fluctuations on the detection plane.
We exert control over the noise level by tuning the camera exposure time. Lower exposure values correspond to larger error amplitudes.
Fixing the CCD settings and analyzing the machine behavior, we map the exposure time into a normalized noise level $\rho \in [0,1]$, which indicates the probability at each iteration to measure a false decrease of the cost function and to erroneously flip a spin.
The minimum noise level available, $\rho=0$, corresponds to faults in the feedback loop coming from spontaneous optical fluctuations in the setup.

\begin{figure*}[t!]
\centering
\vspace*{-0.5cm}
\hspace*{-0.3cm}
\includegraphics[width=2.1\columnwidth]{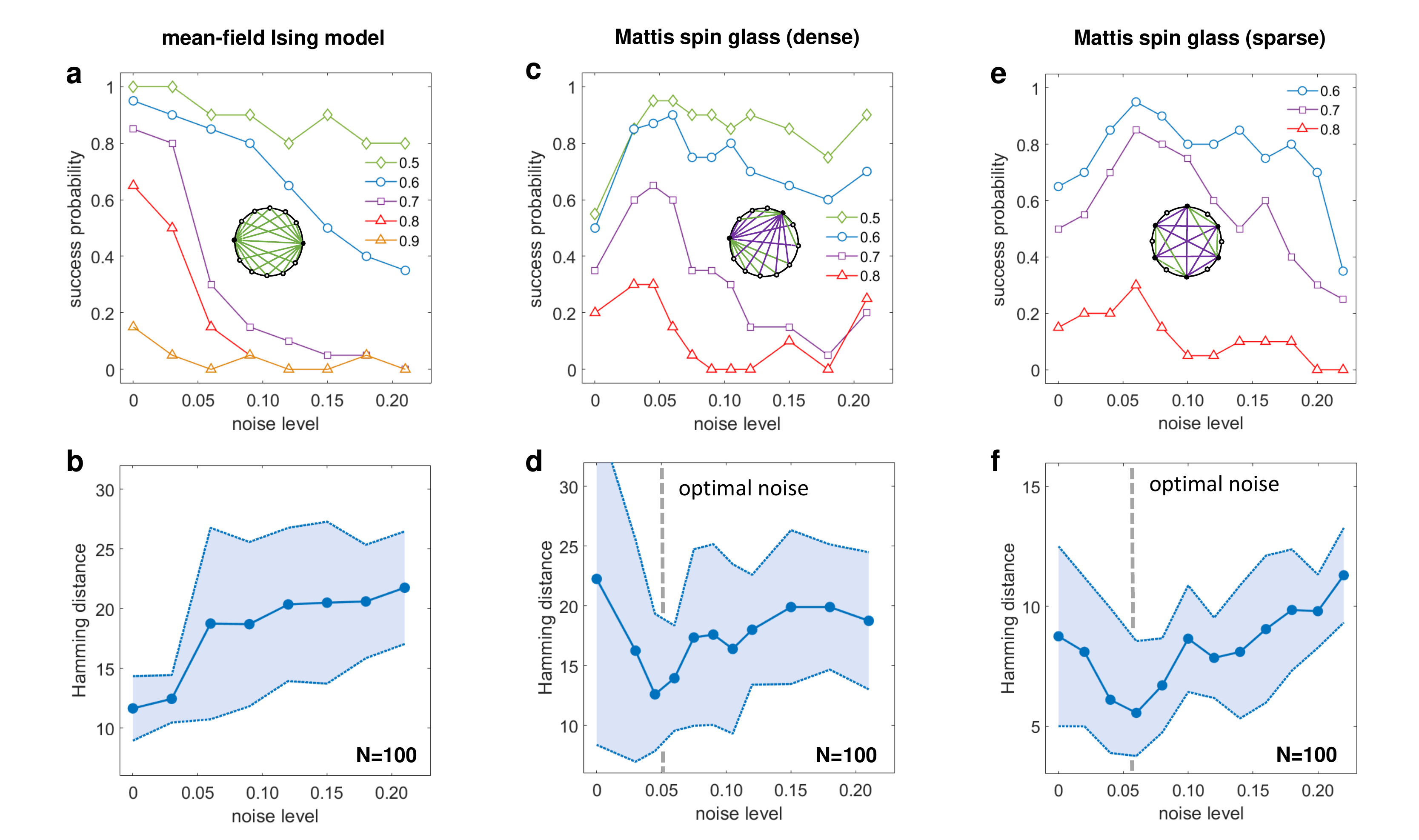} 
\vspace*{-0.5cm}
\caption{{\bf Optimal noise level in spatial-photonic Ising machines.} (a,c,e)  Success probability and (b,d,f) mean Hamming distance varying the noise level for various $N=100$ Ising models: (a-b)  mean-field Ising model, (c-d) dense and (e-f) sparse Mattis spin glasses. The problem graphs are inset. 
Different colours in (a,c,e) indicate data obtained at the specified accuracy level. Shaded regions indicate statistical error intervals.
The existence of an optimal noise level for frustrated models is signalled by a minimum in the Hamming distance and a corresponding maximum in the success probability [dotted line in (d) and (f)].
}
\vspace{-0.1cm}
\label{Figure3}
\end{figure*}

\subsection{Optimization with spontaneous noise}

We first quantify the solutions found by our Ising machine for $\rho=0$, a condition analogous to previously reported experiments \cite{Pierangeli2019}.
We optically implement two different classes of Ising systems: mean-field and Mattis spin glasses.
In a mean-field Ising model, also known as infinite-range Ising model, the spins are all-to-all coupled with the same positive interaction strength. 
This problem is realized using a plane wave of constant amplitude $\xi_i=E_0$ and maximizing the intensity detected on a single CCD mode, 
so that $\tilde{I}_T=c$, being $c$ an arbitrary constant, and $\bar{J}=c E_0^2 $.
In Mattis spin glasses the pairwise interaction is given as a product of two independent variables $J_{ij} \propto \xi_i \xi_j$ \cite{Mattis1976, Nishimori2001}. In this case, since the couplings are both positive and negative, a minimal amount of noise introduces frustration \cite{Nishimori2001}. 
In our spatial optical setting, pairs of negatively-coupled spins correspond to points of the optical field that give destructive interference on a fixed CCD mode [see inset in Fig. 2(d)]. Sparsity is implemented all-optically by decoupling a random subset of spin via amplitude modulation (i.e., $\xi_i=0$ for blocks of spins). Since the interaction matrix is given in any case as a product of separate amplitude values, the sparse Ising Hamiltonian still maintains the form of a Mattis model.

Optical ground states found by the device for a mean-field system of $N=100$ spins are reported in Fig.~2(a-c). The evolution of a spin configuration
initialized to a random distribution is shown in Fig.~2(b) for a single machine run; after thousand iterations, a low-temperature ferromagnetic state is measured. The energy probability distribution of these ground states is reported in Fig.~2(c); a peak close to the known minimum energy value [red line in Fig.~2(c)] indicates that ground states of the Ising Hamiltonian are successfully found. 
In Fig.~2(d) we show the graph for a Mattis spin glass instance along with the corresponding target image. In agreement with the programmed Hamiltonian \cite{Nishimori2001}, the spin evolution in Fig.~2(e) exhibits the formation of two domains with opposite magnetization and equal size.
However, from the ground-state energies in Fig.~2(f), we observe a reduced ability of the machine in solving frustrated models.
As we demonstrate hereafter, increasing the noise level is an effective way to improve the machine performance on problems affected by trapping in local energy minima.

\section{Effect of the noise level on the Ising machine performance}

To introduce controllable noise, we exploits the leaky detection process previously described. 
The Ising machine is made to operate under different noise levels.
To quantify the performance when the setup is initialized to different parameters, we use two distinct and complementary quantities: the success probability $p_s$ and the Hamming distance $h$.  To evaluate the success of the computation, we consider the correlation between the measured spin configuration and the known optimal solution, being $C=\pm 1$ for the zero-temperature spin system in the lowest energy state \cite{Nishimori2001}.
A machine run is defined as successful if its final state gives $|C|> a$, being $a$ any fixed accuracy; the success probability $p_s$ is the fraction of repeated experiments with random initial conditions that converged successfully. 
The Hamming distance is a metric used in information theory. Here, $h$ indicates the number of units (spins) that need to be inverted to reach the minimum energy configuration \cite{Boixo2014}. We remark that using this quantity to characterize the quality of Ising machines is much more accurate than the ground-state energy.  In fact, spin configurations very far from the known solution can still have energies comparable with the optimal one.

Figure 3 illustrates the performance of the spatial-photonic Ising machine as we vary the noise level. For the infinite-range Ising model, we found a success probability that decays as noise increases, a behavior independent of the selected accuracy [Fig.~3(a)]. The measured $h$ in Fig.~3(b) has a growing trend, thus indicating that, on these models, the best performance takes place at minimum noise. Noise reduces the effectiveness of the minimization. A completely different picture emerges when solving frustrated Ising models.
As shown in Fig.~3(c) for dense Mattis spin glass instances, the low success probability observed under spontaneous noise rapidly increases as additional fluctuations are introduced.  $p_s$ has a maximum value and then decays to zero for large noise levels. Moreover, the mean Hamming distance as a function of the noise level presents an evident minimum [Fig.~3(d)], which indicates an optimal noise level, $\rho_{opt}\approx0.05$, that enhances the device efficiency.
A similar behavior is found on sparse Mattis spin glasses, as reported in Fig.~3(e-f), with $\rho_{opt}\approx0.07$.
This demonstrates that an optimal noise level promotes the exploration of energy landscapes with many minima during the photonic optimization.

\subsection{Scaling of the optimal noise level}

We investigate how the noise-enhanced machine operation depends on the system size. While keeping constant the SLM active area, we vary the total number of spins and, for each system size $N$, we perform the experiments at different noise levels.
The results obtained on dense Mattis spin glasses are shown in Fig.~4.
We observe an optimal noise level that significantly depends on the spin number, with values that grow and saturate as the system size increases.
For small-scale systems ($N=16$) additional noise yields only limited advantages due to finite size effects and
the small number of frustrated configurations. Differently, a constant optimal level $\rho_{opt}\approx0.07$ enhances the optimization on large scales. This specific value is not a general recipe to improve the Ising machine, but guarantees improved performance on specific problem graphs in our setup. 
Ising instances with distinct properties (graph type, connection density etc.) would have diverse optimal levels, having landscapes with different features.
These findings establish the noise level as a hyperparameter of the photonic computing device.

Another important fact is that residual errors do not increase rapidly with $N$.  To prove this property, Figure ~4 shows the Hamming fraction $h/N$
(i.e., the mean Hamming distance normalized to the spin number) as a function of $N$. Residual errors shows a sublinear increase with minor oscillations as $N$ varies. This preliminary evidence suggest a smooth dependence of the machine accuracy on the system size, although larger systems naturally require more machine iterations to converge to their ground state.
According to thermodynamic considerations, this is consistent with the presence of an effective thermal bath for the spin state. It suggest that lowering the effective temperature of the machine by improving the experimental setup can lead to larger ground-state probabilities even for large-scale models.

\begin{figure}[t!]
\centering
\vspace*{-0.3cm}
\hspace*{-0.6cm}
\includegraphics[width=1.1\columnwidth]{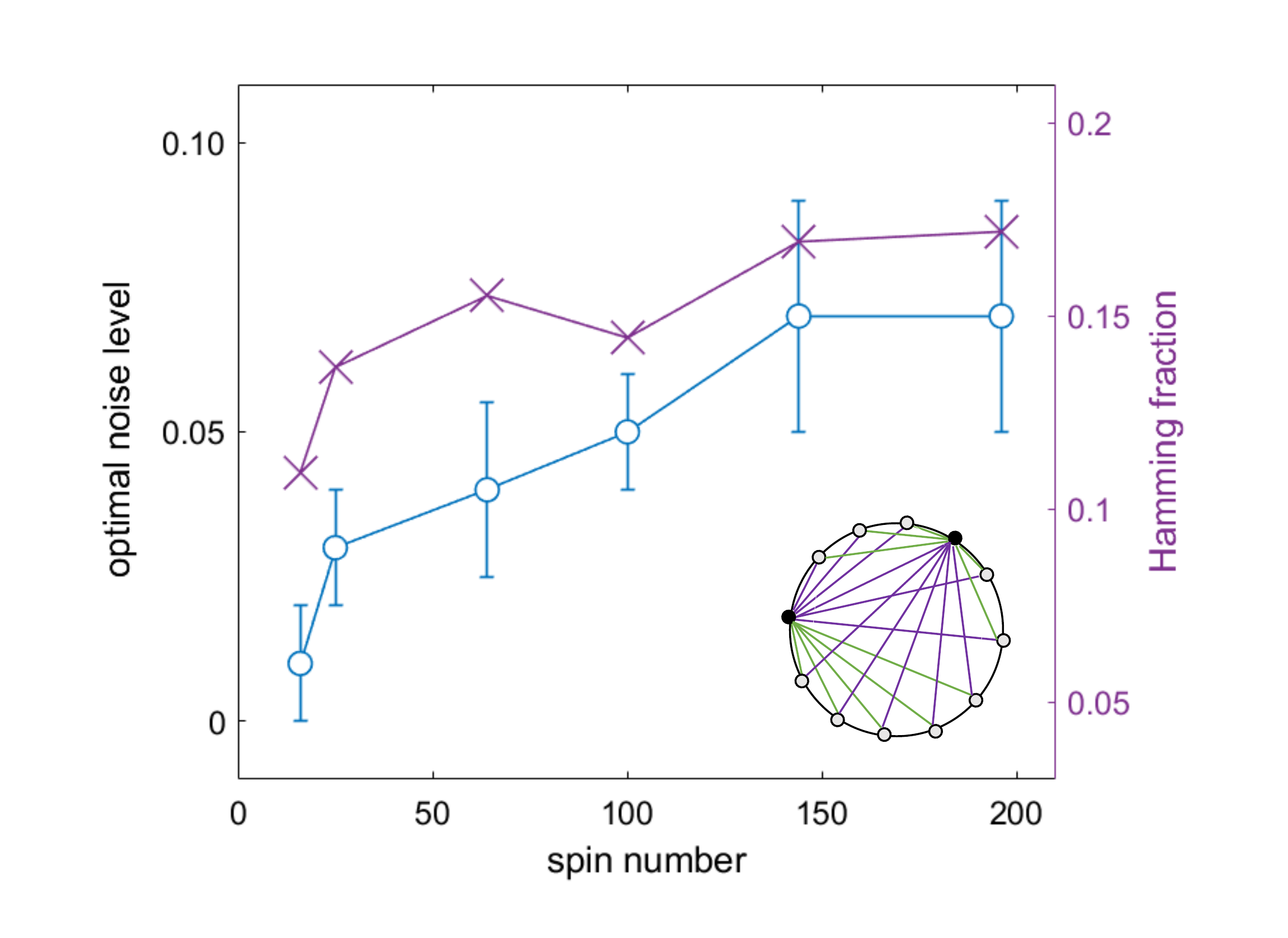} 
\vspace*{-0.7cm}
\caption{ {\bf Scaling properties.} Optimal noise level (blue circles) as a function of the number of spins for dense Mattis spin glasses (inset graph). 
Purple dots indicate the measured Hamming fraction ($h/N$) for various system sizes.
}
\vspace{-0.1cm}
\label{Figure4}
\end{figure}

\section{Conclusions}

Understanding the role of noise on optical Ising machines and neuromorphic devices is crucial for their application on large-scale computational tasks. 
In particular, noise-tolerant settings are attractive candidates for developing unconventional computing architectures. 
We have reported the first evidence that spatial-photonic Ising machines can take advantage of noise in solving large-scale hard optimization problems.
Devices based on spatial light modulation are scalable to larger sizes and can potentially host systems consisting of several thousands of spins. 
In particular, our computing setting can exploit the potential of nanophotonic light-modulation devices. 
Tunable dielectric metasurfaces, which allow to control both phase and polarization of the optical wavefront with subwavelength spatial resolution \cite{Kamalia2018}, can act as high-density phase modulators, enabling the integration of SLM-based Ising machines on a photonic chip. 
For example, more than $10^6$ optical spins over square millimeter could be obtained through the development of novel SLM technologies that integrate nanoantennas into liquid crystal cells \cite{Li2019}. Alternative nanophotonic platforms that employ electro-optic microcavity arrays would allow to achieve high fill factors along with phase-only modulation at GHz speeds \cite{Peng2019}. At present, the iteration time of our Ising machine can be reduced to a few milliseconds by exploiting the most recent microelectromechanical SLM technologies \cite{Blochet2017, Tzang2019}. 
Spatial-photonic Ising machines are thus a promising approach for large-scale ultra-fast optical computing.

In conclusion, introducing a noisy-feedback mechanism in a SLM-based scheme, we have demonstrated the existence of an optimal noise level 
enhancing the machine performance on frustrated Ising models.
Noise can hence be exploited as a tunable parameter to improve the exploration of energy landscapes with many minima, 
an interesting property that has been identified also in neural-network-based nanophotonic Ising samplers \cite{Roques-Carmes2020, Prabhu2019}.
Photonic Ising machines with controllable noise represent a route to realize photonic simulations of phase transition and finite-temperature phenomena.
Our results show that noise is a valuable resource for optical computing, opening important possibilities for realizing classical and quantum annealing.

\vspace*{0.2cm}
We acknowledge funding from Sapienza Ateneo, SAPIExcellence 2019 (SPIM project), QuantERA ERA-NET Co-fund (Grant No. 731473, project QUOMPLEX), PRIN PELM 2017 and H2020 PhoQus project (Grant No. 820392).
We thank Mr. MD Deen Islam for technical support in the laboratory.

\end{document}